\documentclass[10pt]{article}

\usepackage{amsmath}
\usepackage{amssymb}

\usepackage{graphicx}

\usepackage{cite}

\usepackage{color} 

\usepackage{setspace} 
\usepackage[labelfont=bf,labelsep=period,justification=raggedright]{caption}

\makeatletter
\renewcommand{\@biblabel}[1]{\quad#1.}
\makeatother

\date{}

\graphicspath{{../Figures/}}

\begin{document}

\begin{flushleft}
{\Large
\textbf{Balancing noise and plasticity in eukaryotic gene expression}
}

Djordje Baji\'{c}$^1$ and Juan F. Poyatos$^{1*}$\\

\bf $^1$ Logic of Genomic Systems Laboratory (CNB-CSIC), Madrid, Spain
\end{flushleft}

\section*{Abstract}



Coupling the control of expression stochasticity (noise) to the ability of expression change (plasticity) can alter gene function and influence adaptation. A number of factors, such as transcription re-initiation, strong chromatin regulation or genome neighboring organization, underlie this coupling. However, these factors do not necessarily combine in equivalent ways and strengths in all genes. Can we identify then alternative architectures that modulate in distinct ways the linkage of noise and plasticity? Here we first show that strong chromatin regulation, commonly viewed as a source of coupling, can lead to plasticity without noise. The nature of this regulation is relevant too, with plastic but noiseless genes being subjected to general activators whereas plastic and noisy genes experience more specific repression. Contrarily, in genes exhibiting poor transcriptional control, it is translational efficiency what separates noise from plasticity, a pattern related to transcript length. This additionally implies that genome neighboring organization --as modifier-- appears only effective in highly plastic genes. In this class, we confirm bidirectional promoters (bipromoters) as a configuration capable to reduce coupling by abating noise but also reveal an important trade-off, since bipromoters also decrease plasticity. This presents ultimately a paradox between intergenic distances and modulation, with short intergenic distances both associated and disassociated to noise at different plasticity levels. Balancing the coupling among different types of expression variability appears as a potential shaping force of genome regulation and organization. This is reflected in the use of different control strategies at genes with different sets of functional constraints.

\newpage
\section*{Introduction}
Variation in gene expression is observed between closely related species, even when the specific gene coding sequence is largely conserved, e.g.,~\cite{Enard2002}. Within a species, expression can fluctuate following a perturbation (environmental or genetic) and even in the absence of perturbations variation among individuals is found -- this being often interpreted as disturbing noise~\cite{Fraser2004}.  What molecular factors determine these fluctuations? Are these factors subjected to selection pressures? And which general trends on expression variability can one identify at the genomic level?\vspace*{12pt}

Partial answers to these questions were recently reached by using high-throughput experiments on the budding yeast {\it Saccharomyces cerevisiae}. Noise was measured in $>$2,500 proteins using GFP-tagged yeast strains~\cite{Newman2006} and this validated the contribution of mRNA dynamics to protein noise. Both protein function (e.g., housekeeping proteins exhibiting low noise, while stress-response proteins being noisy) and chromatin dynamics (transitions between active/inactive states) were also shown to correlate with noise. Moreover, expression plasticity (responsiveness of {\it S. cerevisiae} genes to change in external conditions) and divergence (among closely related species) were also quantified with the use of a compendium of genome-wide expression profiles in four yeasts~\cite{Tirosh2006}. Genes presenting a TATA box in their promoter showed higher interspecies variability, controlling for function, which suggested the influence of transcription re-initiation mechanisms and bursting expression~\cite{Blake2003}. Similarly, response to mutations (using mutation-accumulation experiments~\cite{Landry2007}) identified TATA boxes and {\it trans}-mutational target sizes (number of proteins influencing the expression of a focal gene) as determinants of neutral variability. Finally, the production of a complete nucleosome occupancy map covering $\sim$81\% of the genome~\cite{Lee2007} helped identify how different (absolute and relative) occupancy levels further controls variability.\vspace*{12pt}

These initial findings are leading to new questions. For instance, are all these aspects of variation (short-term --noise/plasticity-- {\it vs.} long-term --divergence) linked to a unifying promoter structure? This is clearly suggested in recent studies, with an emphasis on the role of chromatin regulation~\cite{Tirosh2008,Choi2008,Choi2009}. This strategy could be positive in terms of the economics of regulation~\cite{Warner1999}, but negative in terms of functional conflicts, e.g., need of bipolarity in genome-wide transcription~\cite{Basehoar2004,Kim2011}, presence of gene classes requiring precise but plastic expression~\cite{Lehner2010}, etc. An additional question is to what extent a demand for variation acts as a central force for the organization of genomes and {\it vice versa}, i.e., whether structural genomic features constrain variation~\cite{Batada2007,Wang2011,Woo2011}.\vspace*{12pt}

Here, we first revisited the influence of chromatin regulation in the linkage of noise and plasticity. We observe that both regulatory strength and character modulates this linkage. Plastic genes exhibiting relatively strong chromatin regulation can appear independent of noise, but an extra increase in plasticity associates plasticity with noise. This association --or the lack of it--is revealed in the type of chromatin control, with a contrast between global and specific regulation. While these patterns indicate transcriptional initiation as fundamental mechanism of modulation (as previously suggested, e.g.,~\cite{Blake2003}), we alternatively find that noise uncouples from plasticity in low-plastic genes due to changes in translational efficiency. These distinct modes are confirmed by the differential influence of genomic neighborhood on coupling depending on plasticity. Interestingly, short intergenic distance and bidirectional promoter architecture can both be related to high and low noise.\vspace*{12pt}

\section*{Results }
\subsection*{Chromatin regulation does not always link plasticity to noise}

TATA boxes and high nucleosomal occupancy at the proximal regions of transcriptional starting sites (TSSs) have been recognized as fundamental promoter features leading to gene expression variability~\cite{Basehoar2004,Tirosh2006,Landry2007}. Both features were shown to couple two specific forms of variability, i.e., expression noise and plasticity. Linkage between noise and plasticity was additionally associated to a highly dynamic chromatin, as quantified by histone exchange rates~\cite{Lehner2010}. However, histone exchange rates do not fully describe the many {\it trans} factors influencing nucleosome dynamics.\vspace*{12pt}

To better understand how such factors determine the noise-plasticity coupling, we used a score that assesses chromatin regulation effects (CRE), i.e., how much the expression of a given gene varies when deleting its {\it trans}-acting chromatin regulators~\cite{Steinfeld2007,Tirosh2008,Choi2008,Choi2009} (Methods). CRE correlated with high plasticity as expected (Spearman's correlation coefficient $\rho = 0.57, p < 10^{-20}, n = 2045$). We then grouped genes in terms of proximal nucleosome occupancy and computed mean plasticity for those genes exhibiting high or low CRE within each group. Notably, chromatin regulation can induce a relatively high level of plasticity independent of nucleosomal occupancy and presence of TATA promoters (Figure~1). Is it possible to identify genes with particularly strong plasticity? This distinguishes genes whose promoters present both high proximal nucleosomal occupancy --increasing sensitivity to regulation (Additional file 1: Figure~S1)~\cite{Choi2009}-- and presence of TATA box --that results in bursting transcription and increased transcriptional efficiency~\cite{Blake2003}-- which in turn involves coupling to noise (confirming earlier reports~\cite{Newman2006,Tirosh2008}, Additional file 1: Figure~S2).\vspace*{12pt}

The crucial effect of the high (proximal) nucleosomal occupancy to enhance coupling is emphasized by the pronounced nucleosome depleted region (NDR) exhibited by a subset of TATA-containing genes with low noise (NDRs are similarly observed in TATAless genes, Additional file 1: Figure S3). Moreover, if the noise-plasticity coupling had its origin in the stability of the transcriptional apparatus at the promoter, this would predict the presence of coupling in TATAless genes with a SAGA-dominated initiation (that also produces transcriptional bursting~\cite{Zenklusen2008}). This is indeed what we observed (noise-plasticity correlation in TATAless and SAGA dominated genes, $\rho=0.51, p=1.2\,{\rm x}\, 10^{-5}, n=66$, see also Additional file 1: Figure~S4).

\subsection*{High plasticity implies different {\it trans-}regulation strategies when coupled/uncoupled to noise}
To further appreciate what determines the coupling (or uncoupling) of noise with plasticity, we inspected potential differences in the type of chromatin regulation. We computed the mean effect in expression of a compendium of mutations in regulators~\cite{Steinfeld2007} (CRE score before represents a subset, see Methods) on plastic genes. This analysis highlighted a strong anti-correlation between the effect of perturbations in low-noise high-plasticity genes (LNHP, see Methods for definition of these classes) and high-noise high-plasticity genes (HNHP, $\rho= -0.83, p<2.2\,$x$\,10^{-16}, n = 170$, Figure~2A; this correlation is much stronger than the expected baseline correlation, Additional file 1: Figure~S5). This confirms mechanistically a complementary program of regulation between these two groups of genes (enriched by growth --ribosomal-- and stress genes, respectively~\cite{Basehoar2004,Lopez-Maury2008}), to be added to the previously observed distinctions in promoter nucleosome occupancy (Additional file 1: Figure~S6), and histone modification enrichment~\cite{Kim2011}.\vspace*{12pt}

More specifically, perturbations affecting LNHP genes commonly cause a decrease in expression (Figure~2A, red dots), i.e., regulators activate expression, and dominantly correspond to general transcription factors  (20 out of 41). Even if most of these perturbations involved mutations in TAF1 --which is a general transcription factor (also associated to chromatin modulation activities, e.g.,~\cite{Huisinga2004,Durant2006}) needed for the expression of nearly 90$\%$ of yeast genes--, its perturbation affected LNHP genes more significantly than low plasticity (LP) ones (Figure~2B; this suggests these genes as preferred targets of TAF1 broad regulatory action, see Additional file 1: Supplement). In contrast, a wider range of different regulators affected HNHP genes, and the majority of these were independent {\it trans}-acting chromatin regulators (92 out of 135) whose deletion activated gene expression (Figure~2A, blue dots), i.e., regulators repress expression (Additional file 2: Table~S1).

We also found that some regulators whose deletion results in decreasing expression level in the majority of LNHP genes tends in comparison to increase it in the majority of HNHP (Figure 3). As expected, this is not related to general transcription factors, whose deletion reduce expression level, but to many specific chromatin regulators and, notably, to histones. As much as 81$\%$ of these histone deletions (see also Additional file 1: Figure~S7) caused an increase in the expression of most HNHP genes, while they decreased the expression level of the majority of LNHP. A recent result can help us understand this~\cite{Kim2011}. Namely, LNHP genes are greatly enriched in activating marks (mostly acetylations), a strong change in acetylation level being observed when repressed. Therefore, acetylated histones are probably essential for the expression of these genes. On the contrary, HNHP genes do not show such changes in histone acetylation status, but they reduce their occupancy level when activated. Histone deletion is in this case more likely to impede the formation of repressive nucleosomes, resulting in a more frequently open promoter and increasing expression level (see Additional file 1: Supplement for more discussion, and Additional File 1: Figure~S8 for a detailed profile of histone modifications).

\subsection*{Noise in low-plasticity genes arises from enhanced translational efficiency}

We noted that LP genes also present differential coupling to noise. In contrast to HP genes, this disparity does not seem to respond to transcriptional-based determinants. LP genes hardly present TATA promoters (5.1$\%$ --26/513-- in LP, 22.4$\%$ --343/1529-- in the rest, $p= 1.6 \,{\rm x}\, 10^{-18}$, $\chi^2$-test), display pronounced NDRs (mean proximal nucleosomal occupancy LP: - 1.70 --$n$ = 513--, rest: -1.46 --$n$ = 1532-- $p = 7.6\,{\rm x}\, 10^{-7}$, Kolmogorov-Smirnov KS-test, see also Methods) and are poorly regulated by chromatin (mean CRE LP: 0.58 --$n$ = 513--, rest: 0.70 --$n$ = 1532--  $p<2.2\,{\rm x}\,10^{-16}$, KS-test). A notable feature of these genes is their enrichment in histone H2A.Z at promoters, which has been already noted~\cite{Kim2011} and is thought to help stabilizing the NDR (with our dataset, mean LP: 0.41, $n$ = 365, rest: 0.12, $n$ = 1144, $p =1.2 \,{\rm x}\, 10^{-9}$, KS-test, see also Additional file 1: Figure~S8). Indeed, there are not observable differences in all these factors when considering low and high noise subgroups within the LP set (data not shown).\vspace*{12pt}

We thus inspected if uncoupling could be associated in these genes to translation as this is known to control noise~\cite{Ozbudak2002,Raser2005}. Our analysis shows that noise in LP genes is correlated with translational efficiency~\cite{MacKay2004} and ribosomal density~\cite{Arava2003} ($\rho=0.22, p=7.9\,{\rm x}\, 10^{-5}$, and $\rho = 0.21, p = 1.6\,{\rm x}\, 10^{-4}$, respectively; $n=327$, and Figure~4A,B) while we did not observe this in highly plastic genes ($\rho= 0.06, p=0.32 $ and $ \rho = 0.08, p = 0.14$, respectively; $n=312$). If translation controls noise in LP genes, then noise should also covariate with factors influencing translation efficiency such as ORF length or codon bias (see, for instance,~\cite{Arava2003,Lackner2007}). In LP genes, translational efficiency correlated more strongly with ORF length ($\rho=-0.58, p=6.7\,{\rm x}\,10^{-31}, n=327$) than with frequency of optimal codons (FOP, $\rho=0.31, p = 9.9 \,{\rm x}\,10^{-9}, n=327$). Consistently, noise correlated with ORF length ($\rho=-0.18, p=5.5\,{\rm x}\,10^{-5}, n=513$ and Figure~4C) but not so with FOP ($\rho=0.07, p=0.12, n=513$). On the other hand, noise correlated with ORF length in an opposite way in HP genes ($\rho = 0.20, p= 5.6 \,{\rm x}\,10^{-6}, n=309$) which probably reflects complementary constraints on gene length (e.g., low noise genes in the HP class are mostly ribosomal genes,  see Additional file 1: Figure~S9. These genes may exhibit short length due to minimization of biosynthetic costs given their high expression~\cite{Warringer2006}).\vspace*{12pt}

\subsection*{Genomic neighborhood modulates the noise-plasticity coupling}

What other mechanisms could modulate the noise-plasticity coupling? The specific architecture of the genomic neighborhood of a gene appears as a possible candidate. Indeed, it was recently found evidence for how bidirectional promoters (bipromoters) could reduce noise by favoring nucleosome depletion~\cite{Wang2011,Woo2011}. We confirm this result for the full dataset (mean noise bipromoter = 0.0796, not bipromoter = 0.0965, Wilcoxon $p=1.2 \,{\rm x}\,10^{-8}$) and validate as well that potentially noise-sensitive gene classes are enriched in bipromoters: essential (465/1062, 43.8$\%$) compared to nonessential (1651/4577, 36.1$\%, p=3.4\,{\rm x}\, 10^{-6}$,  Fisher's Exact test) and genes coding for protein complex subunits (666/1565, 42.6$\%$) compared to the rest (1486/4195, 35.4$\%, p=8.1 \,{\rm x}\, 10^{-7}$,  Fisher's Exact test).\vspace*{12pt}

However, as we have shown, noise in low-plasticity genes is modulated mostly at the translational level, and consequently should not be affected by the presence of bipromoters. Although (nucleosome) depletion is observed independently of plasticity, it does not seem to affect noise in the LP class. In contrast, the noise-reduction effect of bipromoters is observed in HP genes, as expected~\cite{Wang2011} (Additional file 3: Table~S2). This is further corroborated by the enrichment of bipromoters observed in LNHP genes (27/66, 41$\%$) compared to HNHP (47/236, 20$\%,\, p=1.0\,{\rm x}\,10^{-3}$, Fisher's Exact test). Consistently, this enrichment is not significant in LNLP genes (75/145, 52$\%$) compared to HNLP (38/67, 57$\%$, Fisher's Exact test, $p = 0.6$). This pattern is confirmed in noise-sensitive genes: while they are enriched in bipromoters in the HP group (79/237, 33.3$\%$, compared to 67/274, 24.5$\%$ in non noise-sensitive, i.e., noise-tolerant, $p=0.03$, Fisher's Exact test), in the rest (low and medium plasticity) there are no observable differences in bipromoter frequency (noise-sensitive 348/745, 46.7$\%$; and noise-tolerant 348/789, 44.1$\%, p=0.33$, Fisher's Exact test). Notably, low and medium plasticity (noise-sensitive) genes do exhibit a difference in noise (noise-sensitive genes conform 257/442, 58.1$\%$ of the LN, but only 109/252, 41.6$\%$ of the HN, $p =2.4\,{\rm x}\, 10^{-5}$, Fisher's Exact test). If not to bipromoters, this difference could be attributed, as we discussed, to differences in ORF length (mean length noise-sensitive: 1732.0, noise-tolerant: 1517.2, $p=1.9\,{\rm x}\, 10^{-4}$, Wilcoxon test).\vspace*{12pt}

Reducing noise by bipromoters could additionally decrease expression plasticity due to their association to short intergenic distance and nucleosome depletion, and this we actually distinguished (Additional file 3: Table~S2). This suggests then a limitation on the adequacy of bipromoters for reducing noise. Interestingly, we detect a strong bipromoter-independent  effect in noise-sensitive HP genes (Additional file 4: Table~S3). We thus hypothesized that there could be a tendency to evolve noise-abating mechanisms that affect plasticity more weakly, thereby uncoupling it from noise. Indeed, if we consider only bipromoter HP genes, we find that noise-sensitive ones tend to be TATAless (80$\%, n=79$) compared to noise-tolerant (61$\%, n=67$, $p=0.017$, Fisher's Exact test). In agreement with this, the noise-sensitive group has significantly lower noise (mean noise in bipromoter noise-sensitive = 0.083, mean noise in bipromoter noise-tolerant = 0.124,  $p= 1.0\,{\rm x}\, 10^{-4}$, Wilcoxon test) while the difference in plasticity is not significant (mean plasticity in bipromoter noise-sensitive = 0.105, mean plasticity in bipromoter noise-tolerant = 0.12,  $p= 0.62$, Wilcoxon test) indicating an effective uncoupling of noise from plasticity. Can we identify further features illustrating that noise-sensitive bipromoter genes tend to maintain plasticity levels? Intergenic distances suggest that this could be the case (234 bp for noise-sensitive bipromoters and 190 bp noise-tolerant bipromoters,  $p$ = 0.019, Wilcoxon test).

\subsubsection*{Noncoding transcripts and modulation}

\vspace*{9pt}
The above can be complementary analyzed if we consider all possible local genomic architectures around a focal gene (Figure~5A), i.e., parallel, divergent and bipromoters with a coding or non-coding transcript as upstream partner (noncoding partners include ``cryptic unstable transcripts'', CUTs,  and ``stable untranslated transcripts'', SUTs, see~\cite{Xu2009} and Methods; bipromoters CUTS were recently associated with low noise~\cite{Wang2011}). We computed the coupling between noise and plasticity for each architecture. Coupling is strong for genes with divergent transcripts (independent of the type of upstream partner) and weak for those with a bipromoter with a coding partner (Figure~5A). This further validates the observed absence of bipromoters in HNHP genes and their enrichment in the other three classes (bipromoters are the most commonly found architecture in LNLP, HNLP and LNHP) where they are associated, of course, to short intergenic distances (Figure~5A, see also Additional file 1: Figure~S10). Interestingly, bipromoters of plastic genes with low noise are the ones with the biggest (relative) intergenic distance (with respect to LNLP and HNLP), which suggests again the requirement of a  minimal distance to locate the regulatory demands associated to enhance plasticity (mean distance bipromoters of LNHP: 252 bp, in the LNLP and HNLP groups: 178bp, $p=1.1\,{\rm x}\, 10^{-3}$, Wilcoxon test). Overall, this emphasizes bipromoters as noise-abating architecture only when noise and plasticity are transcriptionally modulated.

\section*{Discussion}
We analyzed the molecular determinants that adjust the linkage between gene expression plasticity and noise in {\it S. cerevisiae}. Noise was confirmed to be connected to plasticity when genes exhibit particular modes of transcription initiation (and re-initiation) related to the presence of TATA boxes at the promoter and strong chromatin regulation~\cite{Blake2003,Basehoar2004,Lee2007,Tirosh2008,Choi2008,Choi2009}. This could suggest a model in which intrinsic noise is a byproduct of the need for plasticity~\cite{Lehner2010}. We show, however, that noisy expression can be observed in genes with low plasticity. These genes are generally simple (poor in transcription factor binding sites and generally TATAless), small and poorly regulated by chromatin effectors, and  they show a prominent nucleosome depleted region (Additional file 1: Figure~S6, see also Figure~5B).\vspace*{12pt}

In this loose regulatory scenario, transcription is likely to be produced by single, isolated in time, initiation events~\cite{Huisinga2004}. For such bacteria-like transcription, noise is anticipated to depend on translational efficiency~\cite{Ozbudak2002,Raser2005} (but see~\cite{Golding2005}) and this is indeed what we notice. In addition, ORF length appears as a strong determinant of ribosome occupancy in this class and thus of translation efficiency (potentially due to the lack of post-transcriptional regulation (see Additional file 1: Figure~S11)~\cite{Lee2011}. Following this model we expect essential genes --usually of low plasticity-- to be large as we observe (size essential genes: 1646 bp, size nonessential: 1468 bp, $p < 0.001$, Wilcoxon-test; see also Figure~4).\vspace*{12pt}

On the other hand, we also establish that relatively high expression plasticity can be reached by chromatin regulation without necessarily coupling it to noise. These genes that are enriched in growth-related functions (such as ribosomal protein genes, RPs) depend strongly on TAF1, an essential subunit of TFIID,  but less strongly on most of the more specific chromatin regulators (as compared to noisy and plastic genes in Figure~2, see further discussion in Additional file 1: Supplement). This indicates that they respond to general, rather than gene specific, regulatory strategies which partly explain the high degree of co-regulation previously observed~\cite{Kim2011}. Beyond this, the group exhibits a characteristic pattern of low nucleosome occupancy in both proximal and distal promoter regions~\cite{Tirosh2008} possibly caused by the strong enrichment in activating histone modifications~\cite{Kim2011}, and particularly acetylations. We hypothesized that rather than a promoter-localized open-chromatin state; these genes could be located at broader open chromatin domains. Indeed, we distinguished that RP genes tend to be localized on broad open-chromatin domains that extend up to at least 40Kb (Additional file 1: Figure~S12), supporting this view.\vspace*{12pt}

The low noise but highly plastic expression is therefore consistent with two (not mutually exclusive) models previously proposed. Firstly, it is consistent with a detailed model~\cite{Raser2004} in which low nucleosome occupancy at the promoter indicates a stable open state, allowing the high expression levels exhibited by these genes (Additional file 3: Figure~S13). The concomitant noise reduction would not be possible if the high expression level would be reached by an increase in transcription or translation efficiency~\cite{Raser2004}. Secondly, at a broader level, the localization in open chromatin, and consequently low noise, genomic regions could also contribute to the small level of noise detected~\cite{Batada2007}.\vspace*{12pt}

Finally, we determined how additional enhancement in plasticity fundamentally associates to strong intrinsic noise. This emphasizes two additional promoter features. Firstly, a particular nucleosome located at the promoter~\cite{Lee2007,Tirosh2008,Choi2008} allows a fine-tuned --and gene-specific-- control of open and close promoter states by external regulators. The repressive effect of this nucleosome is evidenced by the increase in expression level upon histone depletion, an increase only observed in these genes (Figure~3). We demonstrate that many specific chromatin regulators act on these genes, in contrast to their low-noise counterpart (Figure~2) that could reflect also in the lack of co-regulation reported~\cite{Kim2011}. Additionally, during the time lapses that the promoter is in open state, the presence of a TATA box allows the pre-initiation complex to stay assembled firing continuous initiation events. This increases the sensitivity to changes --in the time the promoter stays in open state produced by chromatin remodeling-- thereby allowing an increase in plasticity. Moreover, and even when repressed, the nucleosome can occasionally be destabilized allowing strong bursts of transcription which result in the observed noise. That coupling is related to an efficient transcription initiation is confirmed by the strong coupling found in SAGA but TATAless genes (Additional file 1: Figure~S3), what confirms the model in~\cite{Raser2004} to a genome-wide scale.\vspace*{12pt}

As the critical promoter-covering nucleosome is probably stabilized by particular DNA properties, such as high bendability~\cite{Tirosh2007b,Choi2008}, this could potentially increase the number of phenotype-affecting mutational targets, which could be in turn the cause of the increased expression divergence in these genes (Additional file 1: Figure~S14). Therefore, linkage between different types of variability is mechanistically a consequence of the sophisticated regulatory strategy involving promoter nucleosomes and TATA boxes.  This regulation also brings higher sensibility to chromatin regulation (leading to plasticity), to stochastic nucleosome fluctuations (leading to noise) and to mutational effects (leading to expression divergence).\vspace*{12pt}

The action of these two distinct strategies to modulate noise and plasticity coupling is further emphasized by the structure of the genomic neighborhood of the focal gene under study. While a seemingly general architecture in which (relative) reduction of intergenic distance and enrichment of bipromoters should indicate noise-plasticity uncoupling, this only applies to transcriptional-based modulation (Figure~5A). Indeed, both small integenic distance and bipromoter can be broadly related to high or low noise in poorly plastic genes, where uncoupling is rather associated to other mechanisms of modulating noise (HNLP genes exhibit higher translational efficiency). Moreover, we observe that genes particularly sensitive to noise (e.g., genes specifying proteins in complexes) can separate noise from the requirement of high plasticity (these genes showed larger intergenic distances as bipromoters).\vspace*{12pt}

In view of the exposed implications, can we speculate about the evolution of bipromoters?  Firstly, the {\it S. cerevisiae} genome is highly gene-dense, averaging one gene each $\sim$2 Kb with a median intergene distance in our dataset (considering noncoding transcripts) of 204 bp. A high bipromoter frequency seems then plausible in absence of selection. In addition, we observed a correlation between intergene distance and plasticity ($\rho=0.19, p < 10^{-43}, n=5102$), stronger when only transcripts with divergently oriented upstream partners are considered ($\rho=0.27, p<10^{-57}, n=3271$; Figure~5A). This relationship probably responds to the need of a greater genomic space to accommodate a more complex regulatory landscape, which is in turn needed to achieve controlled expression variability. We observed a strong bias in bipromoter frequency towards genes with low plasticity (Additional file 1: Figure~S15), indicating that these regulatory needs are a major force determining the {\it absence} of bipromoters. In contrast, we suggest that their presence could have an almost neutral origin, as it displays an expectable distribution in view of intergenic distances (Additional file 1: Figure~S16).

\section*{Conclusions}

In sum, the results reported here reveal that transcriptional- and translational-based regulatory strategies are alternatively used to modulate the balance between noise and plasticity in eukaryotic gene expression (Figure~5). These strategies appear clearly associated to distinctive functional (e.g., growth/stress programs in {\it S. cerevisiae}~\cite{Basehoar2004,Lopez-Maury2008,Kim2011}), and genomic constrains (e.g., presence of bipromoters, non-coding transcription or length of coding sequence). Future analysis of additional questions, e.g., role of post-transcriptional regulation (Additional file 1: Figure~S11), potential presence of condition-dependent variation~\cite{Lee2011}, or level and relevance of coupling in higher eukaryotes, should ultimately expose the many aspects of gene expression variation and its evolution.

\section*{Methods}
\subsection*{Gene expression variability}
To quantify gene expression plasticity, i.e. responsiveness to environmental change, the variability in mRNA levels among $>$1500 different growth conditions was measured. as the sum of squares of the $\log_2$-ratios over all these conditions~\cite{Tirosh2006}. Noise, or stochastic variability, was measured by proteomic analysis~\cite{Newman2006}. We used the ``distance to the median''(DM) score, which rules out confounding effects of protein abundance, allowing protein-specific noise levels to be compared. Evolutionary divergence in gene expression was measured as the variation of gene expression between orthologs in in four related yeast species and 32 different conditions~\cite{Tirosh2006}. These three gene expression variability measures we scaled between 0 and 1. After scaling, mean values are 0.062, 0.089, and 0.186 for plasticity, noise and expression divergence, respectively. Moreover, for the noise and plasticity measures, we define three categories ``high'',``medium''and``low'', using percentiles 25 and 75 in each case as boundaries. Thus we obtain groups of genes with high noise and high plasticity (HNHP), high noise and low plasticity (HNLP) and so on (see also Additional file 5: Table~S4). Finally, mRNA level in rich media was obtained from~\cite{Holstege1998} (mean=3.915 mRNA copies/cell).

\subsection*{Genomic localization and neighborhood}
The coordinates of each transcript (coding ORFs--ORF-T; and noncoding, which can be in turn ``cryptic unstable transcripts''--CUTs, and ``stable untranslated transcripts''--SUTs) were obtained from a high resolution transcriptomic analysis~\cite{Xu2009}. For each of these transcripts, we used chromosomic coordinates of transcription start sites (TSS) and transcription end sites (TES), and orientation (strand). This data allowed us to characterize the genomic neighborhood of each transcript, in terms of distance to its upstream partner in bp, orientation of this upstream partner (which can be divergent or parallel). As well, from~\cite{Xu2009} we obtained data describing for each gene whether it is transcribed from a bi-directional promoter based on the existence of a shared nucleosome depleted region (NDR). For the genomic neighborhood analysis, in order to maximize its reliability, we removed from the dataset gene whose upstream partner was a ``pseudogene'' or a ``dubious ORF'', as well as a few confounding cases where adjacent transcripts were overlapping. For some genes upstream distance could not be calculated as TSS and/or TES coordinates could not be accurately determined in the original source (see~\cite{Xu2009}).\vspace*{-9pt}

\subsection*{Promoter characterization and regulation}
The presence/absence of TATA boxes at the promoters was obtained from~\cite{Basehoar2004}. Nucleosome occupancy data for the whole genome was obtained by DNA digestion with micrococcal nuclease and identification of nucleosome-protected fragments by high resolution microarray analysis~\cite{Lee2007}. We use the $\log_2$-ratios provided in the reference. As suggested in~\cite{Tirosh2008}, we obtained two different nucleosome occupancy values for each promoter. Taking as reference the TSS, proximal nucleosome occupancy was the average in the -100 to 0bp region, while distal nucleosome occupancy corresponds to the -400 to -150 bp region. For an idea, the highest occupancy for a proximal region in our dataset was 0.27, and the lowest -3.64.\vspace*{-9pt}

\subsection*{Transcription regulation data}
To explore chromatin regulation, we used a compendium, assembled in~\cite{Steinfeld2007}, consisting of 170 expression profiles for chromatin regulation related mutations (expressed in $\log_2$-ratios). We classified these mutations in three classes (see Additional file 2: Table~S1).``Chromatin'' tag was assigned to mutations in histone acetyltransferases, deacetyltransferases, methylases, demethylases, ubiquitinating and deubiquitinating enzymes, chromatin remodellers and silencing factors. ``General'' tag was assigned to genotypes involving at least one mutation in essential, general transcription factors (TAF1).``Histone'' tag was assigned to mutations in the very histones. As suggested in~\cite{Choi2008}, we normalized each dataset from the compendium to unit variance. The absolute value of the normalized $\log_2$ ratios represented responsiveness measures; the mean responsiveness of each gene represented its ``chromatin regulation effect'' (CRE) or ``histone regulation effect'' (HRE).We used also this normalized dataset without taking the absolute value to analyze the sense of the observed regulation. Data for nucleosome-normalized, chromatin modification states at promoter were obtained from ChromatinDB (http://www.bioinformatics2.wsu.edu/chromatindb) which unifies several experimental genome-wide datasets measuring levels of different histone modifications. For dependence of each gene on general TFs, we used categorical data from~\cite{Huisinga2004} defining for the expression of each gene if it is dominated by TFIID or SAGA complex.\vspace*{-9pt}

\subsection*{Translation related measures}
We used a measure of translation efficiency obtained in~\cite{MacKay2004} and based for each gene in percent of each transcript in polysomes, its ribosome density, and the relative transcript level (mean=4.35, sd=1.72). We used an additional dataset of ribosome density obtained from~\cite{Arava2003} (mean=0.53, sd=0.31).\vspace*{-9pt}

\subsection*{Noise-sensitive genes} We considered essential genes from the {\it Saccharomyces} Genome Deletion Project and genes specifying proteins in complexes~\cite{Wang2009}. Due to a big reduction in sample size, we excluded from this group haploinsufficient genes~\cite{Deutschbauer2005}. However, note that they are virtually not excluded, since 44/46 (95$\%$) of the identified haploinsufficient genes are labeled as either essential or as complex-forming; indeed, 100$\%$ of the haploinsufficient genes located in the HP group are so.


\section*{Authors' contributions}
    JFP conceived the project, DB conducted computational experiments, DB and JFP analyzed experiments and wrote the manuscript. Both authors read and approved the final manuscript.

\section*{Acknowledgements}
  We thank Laurence D. Hurst for comments on an earlier version of the manuscript. This research was partially supported by the Spanish Ministerio de Econom\'{\i}a y Competitividad BFU2011-24691 grant (JFP).

\section*{Competing interests}
The authors declare that they have no competing interests.


\begin{figure}
\centerline{\includegraphics[angle=-90,width=.95\textwidth]{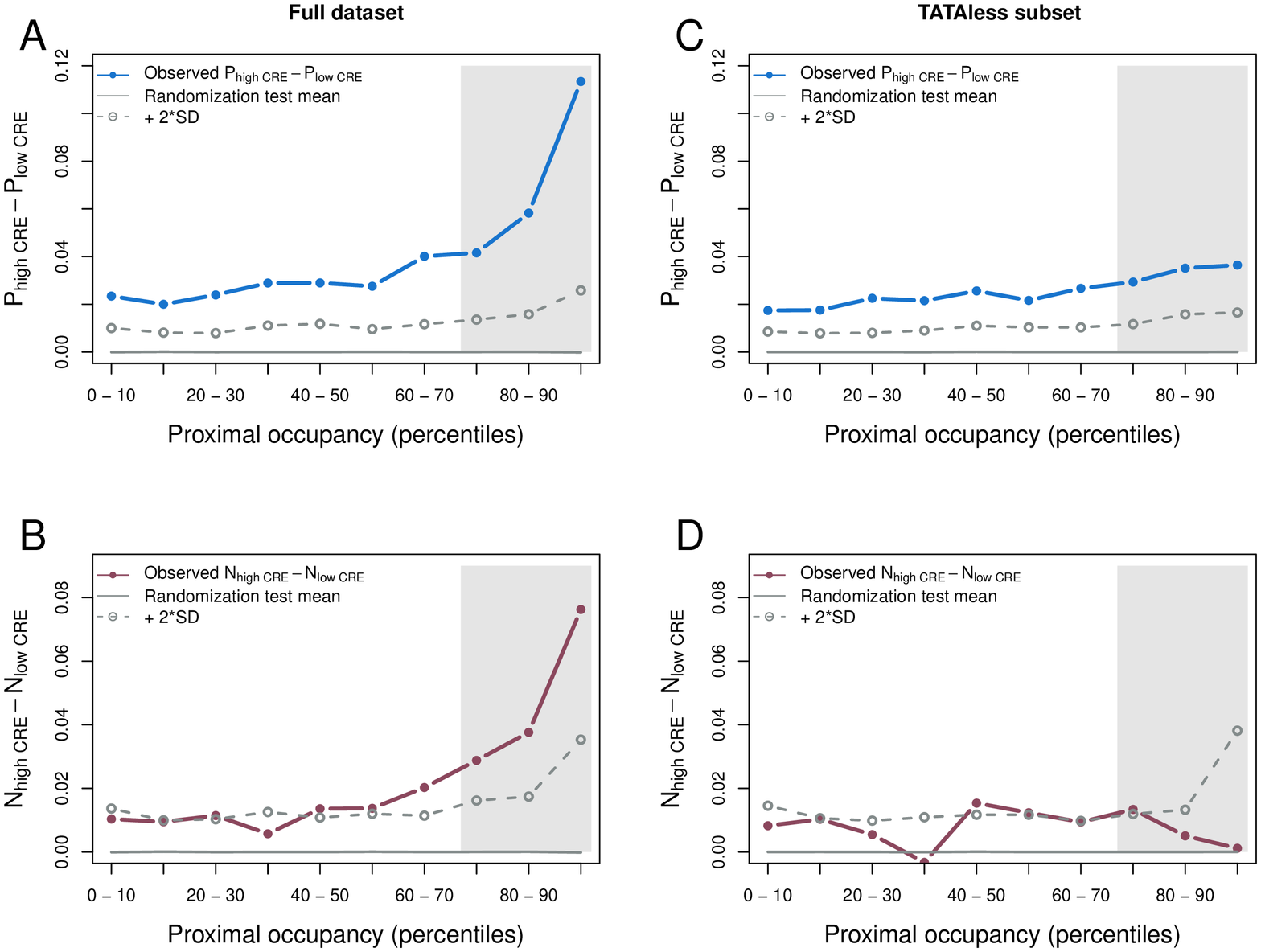}}
 \caption{{Proximal nucleosomal occupancy, chromatin regulation, and the noise-plasticity coupling.}
  We divided the dataset ($n$=2045) in ten equally sized bins of increasing proximal nucleosome occupancy. In each bin, we computed the median chromatin regulation effect (CRE). We plotted the difference in plasticity (\textbf{A}, blue curve) or noise (\textbf{B}, red curve) of genes above/below this median and contrasted the observed values with those expected randomly (permutation test in each bin to depict significance, shown as the mean --gray curve-- and mean plus two standard deviations --dashed gray curves-- obtained with 10000 randomizations). Plasticity is always enhanced by strong chromatin regulation; however, regulation enhances noise only in promoters with high proximal nucleosome occupancy and TATA box (shaded area in A,B). An identical analysis is shown in (\textbf{C}) and (\textbf{D}), but excluding genes with TATA-containing promoters. High occupancy does not lead to increased plasticity/noise in this case (shaded area in C,D).
}
\label{Fig1}
\end{figure}

\begin{figure}
\centerline{\includegraphics[angle=-90,width=.95\textwidth]{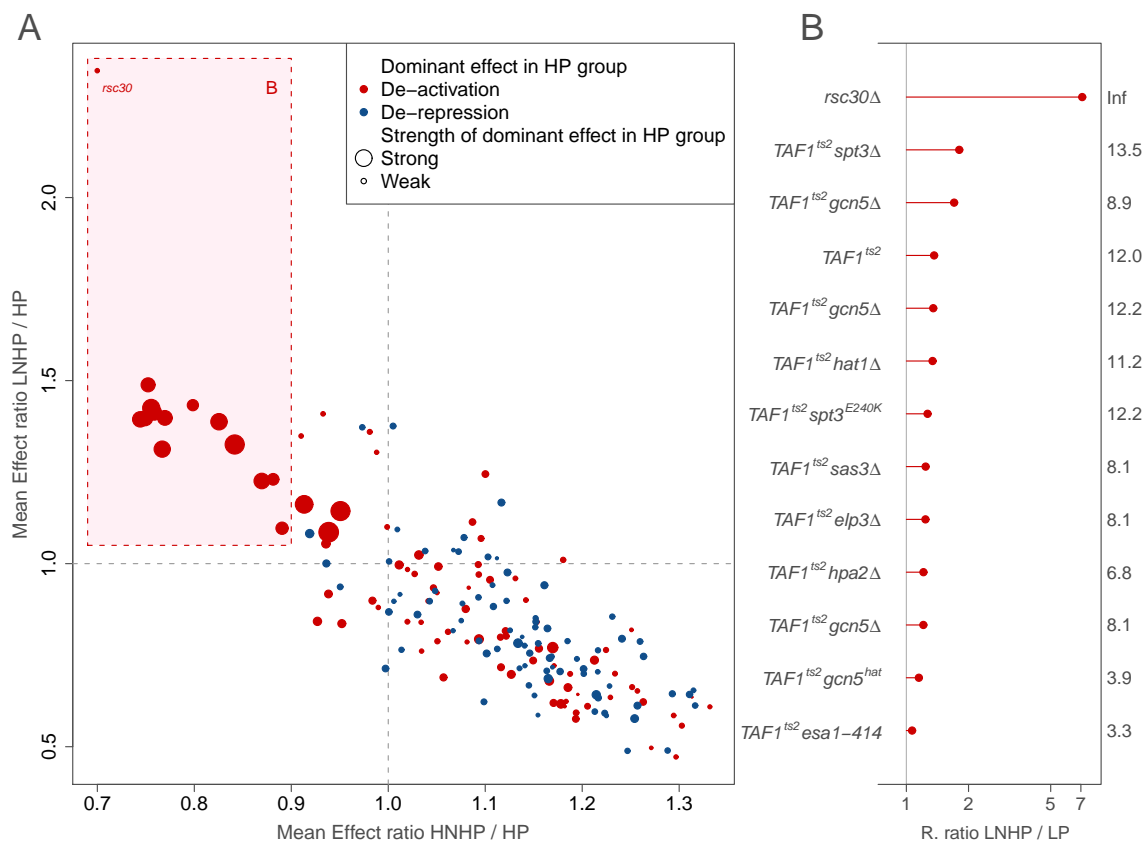}}
 \caption{{Distinct chromatin regulation strategies to achieve noisy or quiet plasticity.} \textbf{A}) Each dot represents the mean effect in the expression of a set of genes in a subclass (HNHP, $x$ coordinate; LNHP, $y$ coordinate; normalized by effect in HP class) when a particular regulator is mutated~\cite{Steinfeld2007}. A ratio $>$1 thus implies that the corresponding subclass is more strongly influenced by certain regulator than the full HP group. A strong negative correlation is found indicating that many regulators are highly specific to either HNHP or LNHP genes. This confirms that these groups are enriched by complementary functional classes (stress and growth related genes, respectively) which are generally regulated in opposite sense~\cite{Basehoar2004,Lopez-Maury2008,Kim2011}. Dot colors denote the dominant effect of the regulator on the HP class (blue; regulator is mostly repressing expression, red; regulator is commonly activating) while sizes describe the strength of the dominant effect; e.g., LNHP genes are frequently affected by strong chromatin activators. \textbf{B}) We examined in detail the effects on LNHP genes (box in A). Except {\it rsc30} (a regulator of ribosomal proteins~\cite{Angus-Hill2001}) all these mutations involved TAF1, which is part of the general transcription factor TFIID~\cite{Huisinga2004,Durant2006}. This essential factor regulates $\sim$90$\%$ of the genes in the genome, not including most of HNHP (which are regulated by SAGA) but including almost all LP genes (see main text). Nevertheless, we observed that all these mutations affected significantly more strongly LNHP than LP genes [K-S tests with FDR-corrected -log($p$-value)'s shown at the right].
}
\label{Fig2}
\end{figure}

\begin{figure}
\centerline{\includegraphics[angle=-90,width=.95\textwidth]{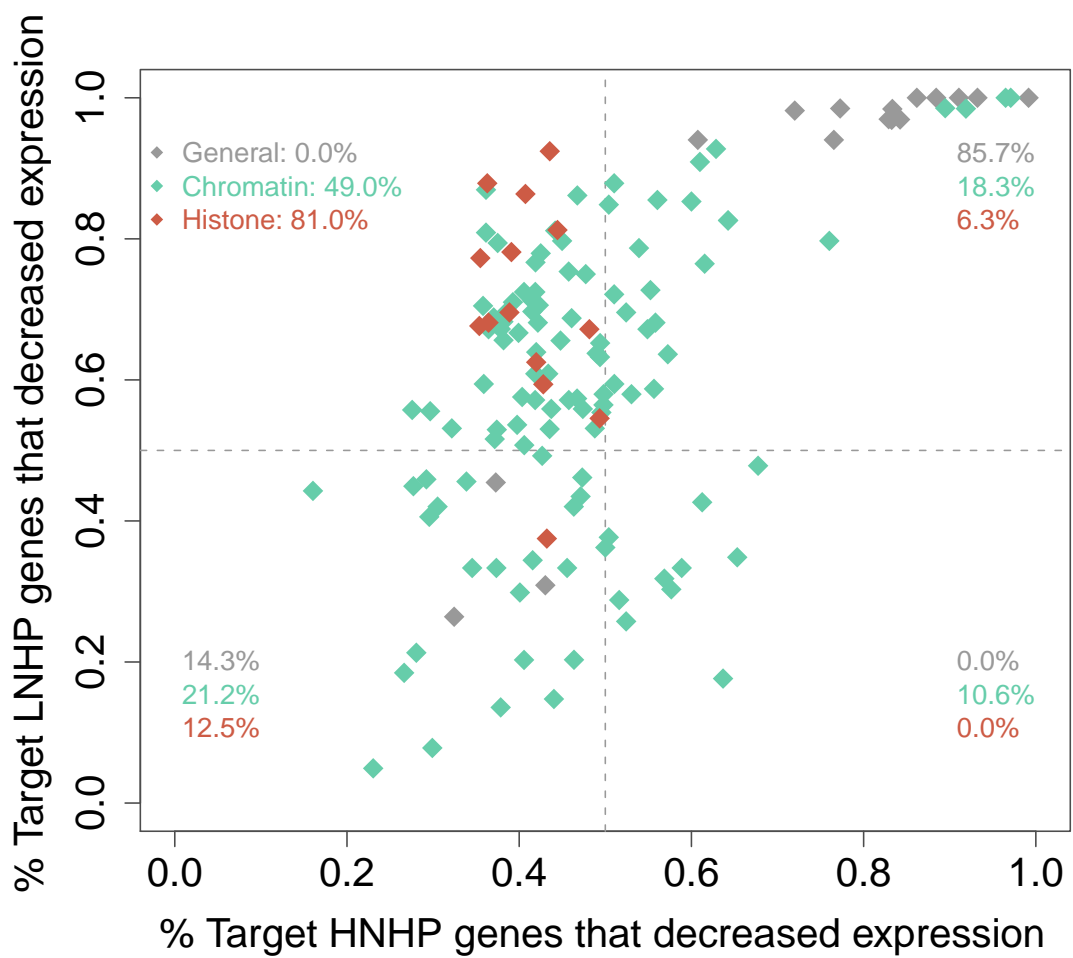}}
\caption{{Dual action of histones.} Both {\it trans}-acting chromatin regulators  and histones tend to have opposite effects in HNHP and LNHP genes. For each mutation from~\cite{Steinfeld2007}, we plot the fraction of genes in the HNHP group and in the LNHP group that decreased expression. Consistently with that observed in Figure~2, we find that mutating as much as $\sim$50$\%$ of chromatin regulators results in the de-activation of the majority of the LNHP genes, but de-repression of most of the HNHP genes. In addition, and perhaps more importantly, we observe that 81$\%$ of mutations in histones also exhibit this behavior. This is probably crucial, and indicates that histones by themselves are needed for repression of HNHP genes and, at the same time, activation of LNHP.
}
\label{Fig3}
\end{figure}


\begin{figure}
\centerline{\includegraphics[angle=-90,width=\textwidth]{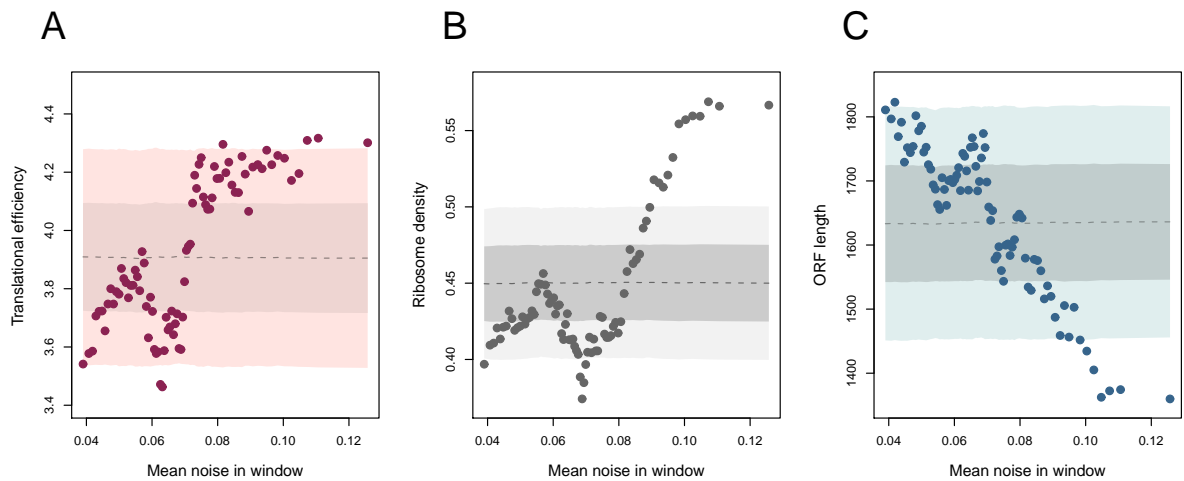}}
\caption{{Noise determinants in low plasticity genes.} Noise in LP genes is related to translational efficiency, which in turn is related to ORF length. We ordered LP genes by increasing noise. We performed a sliding window analysis of translational efficiency (\textbf{A}), ribosomal density (\textbf{B}) and ORF length (\textbf{C}). Shaded regions represent the mean and two standard deviations at each point obtained with the same sliding window analysis over randomized data; the process was repeated 10000 times. See also main text.
}
\label{Fig4}
\end{figure}

\begin{figure}
\centerline{\includegraphics[width=.9\textwidth]{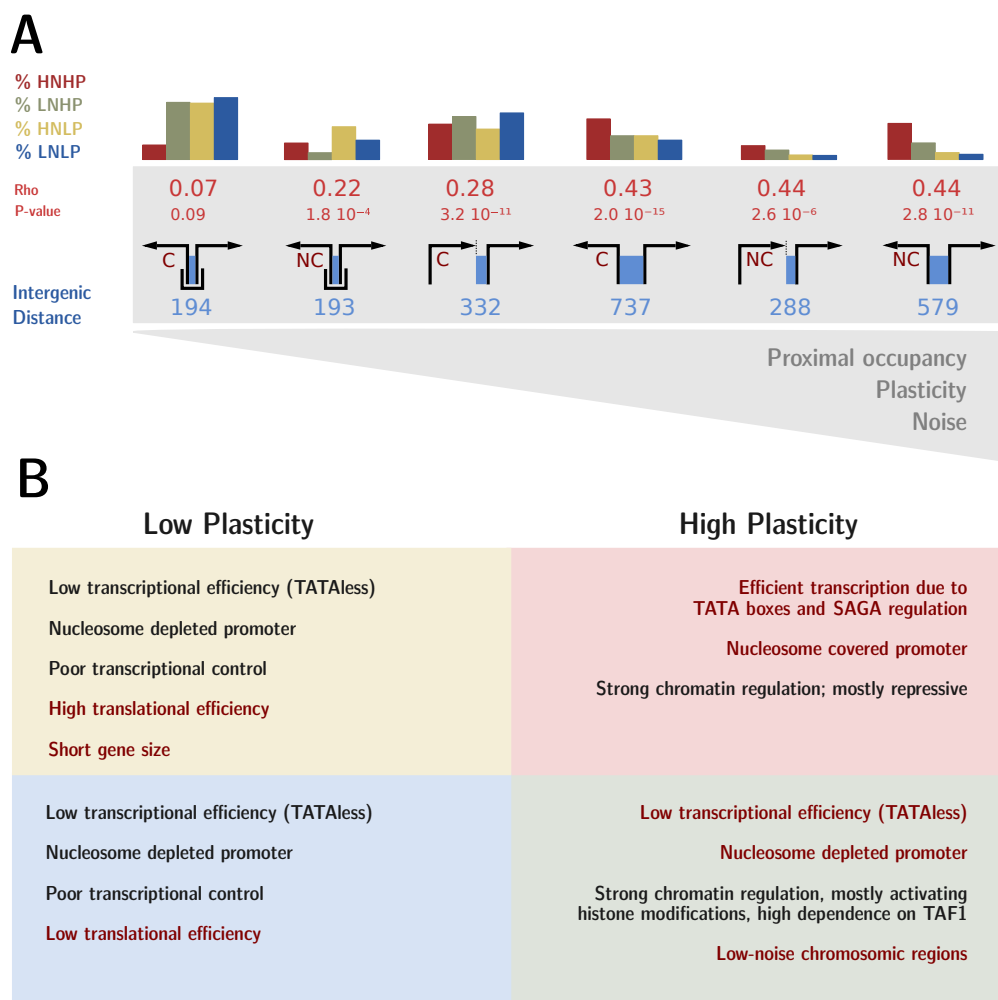}}
\caption{{Noise plasticity coupling is modulated by genomic neighborhood and distinguishes four control strategies overall.} \textbf{A}). A cartoon depicting the different genomic structures (bipromoter, parallel, divergent) upstream of coding genes is shown in ascending order of proximal nucleosome occupancy, plasticity and noise (which coincide). For each structure, we show the average intergenic distance in blue. In red is shown the Spearman $\rho$ coefficient for the observed noise-plasticity correlation. We also show the percent within each class of a given upstream structure , e.g., HNHP mostly exhibit parallel/divergent coding (\textbf{C}) and divergent non-coding (NC) transcripts. \textbf{B}) Four regulatory strategies broadly adjust the noise-plasticity coupling. These strategies emphasize the alternative transcriptional- or translational-based modes of balancing noise and plasticity in yeast.
}
\label{Fig5}
\end{figure}

\end{document}